\documentclass{emulateapj}
\usepackage{graphicx}
\usepackage{gensymb}
\usepackage{multirow}
\usepackage{array}
\usepackage{afterpage}
\usepackage{amsmath}
\usepackage{subfigure}
\usepackage[normalem]{ulem}
\usepackage[flushleft]{threeparttable}
\usepackage{url}
\usepackage[colorlinks=true,citecolor=blue ]{hyperref}

\shorttitle{Supernova Enrichment of Abell 3112 out to R$_{200}$}
\shortauthors{Ezer et al.}

\begin{document}

\title{Uniform Contribution of Supernova Explosions to the Chemical Enrichment of Abell 3112 out to R$_{200}$}

\author{Cemile~Ezer\altaffilmark{1,2}, Esra~Bulbul\altaffilmark{3}, E.~Nihal~Ercan\altaffilmark{1},  Randall~K.~Smith\altaffilmark{2}, Mark~W.~Bautz\altaffilmark{3}, Mike~Loewenstein\altaffilmark{4,5}, Mike~McDonald\altaffilmark{3}, and Eric~D.~Miller\altaffilmark{3}}

\email{cemile.ezer@boun.edu.tr\\}

\altaffiltext{1}{Department of Physics, Bo\u{g}azi\c{c}i University, Istanbul, Turkey\\}
\altaffiltext{2}{Harvard-Smithsonian Center for Astrophysics, 60 Garden Street, Cambridge, MA~02138, USA\\}
\altaffiltext{3}{Kavli Institute for Astrophysics \& Space Research, Massachusetts Institute of Technology, 77 Massachusetts Ave, Cambridge, MA 02139, USA\\}
\altaffiltext{4}{NASA Goddard Space Flight Center, Greenbelt, MD, USA\\}
\altaffiltext{5}{CRESST and X-ray Astrophysics Laboratory, NASA Goddard Space Flight Center, Greenbelt, MD 20771, USA\\}

\begin{abstract}
The spatial distribution of the metals residing in the intra-cluster medium (ICM) of galaxy clusters records all the information on a cluster's nucleosynthesis and chemical enrichment history. We present measurements from a total of 1.2~Ms {\it Suzaku} XIS and 72~ks {\it Chandra} observations of the cool-core galaxy cluster Abell 3112 out its virial radius ($\sim$ 1470 kpc). We find that the ratio of the observed supernova type Ia explosions to the total supernova explosions has a uniform distribution at a level of 12--16\% out to the cluster's virial radius. The observed fraction of type Ia supernova explosions is in agreement with the corresponding fraction found in our Galaxy and the chemical enrichment of our Galaxy.  The non-varying supernova enrichment suggests that the ICM in cluster outskirts was enriched by metals at an early stage before the cluster itself was formed during the period of intense star formation activity. Additionally, we find that the 2D delayed detonation model CDDT produces significantly worse fits to the X-ray spectra compared to simple 1D W7 models. This is due to the relative overestimate of Si, and underestimate of Mg, in these models with respect to the measured abundances.
\end{abstract}


\newpage
\section{Introduction}

Clusters of galaxies are the largest concentrations of confined matter in the Universe. Their deep potential well retains all metals produced by stars and galaxies within the intra-cluster medium (ICM). Improved measurements of the ICM metallicity from X-ray observations provide direct information for the chemical enrichment history of the cluster,  which mainly originates from supernova explosions (SNe) in the stellar populations. Understanding the evolution of the observed cluster enrichment is of vital importance since these structures are unique probes of the nucleosynthesis and chemical enrichment of the Universe.  

X-ray spectra of the ICM contain emission lines of heavy elements, which can only be produced by the late evolutionary stage of stars. From the observational results, the enriched abundance in the ICM is found to be larger than the total metal abundances found in the stellar population within the cluster \citep{portinari04,loewen06}. This implies that the gas is not purely in the primordial state, but a considerable amount of it has been reprocessed within the galaxies and injected into the ICM. \textit{ASCA} observations provided the first measurements of spatial distributions of heavy element abundances, e.g., iron (Fe) and silicon (Si), in clusters of galaxies \citep{baum05}. This pioneering result triggered studies for testing supernova models based on measured supernova (SN) yields. Using the limited \textit{ASCA} measurements of abundance ratios, several studies investigated the efficiency of type Ia (SN Ia) and core collapse  supernova (SN cc) enrichment in the ICM \citep[e.g.,][]{mushotzky97, ishimaru97, finoguenov00, dupke00}. These earlier studies suggest an early homogeneous enrichment by SN cc shortly after the cluster formation with its products well-mixed throughout the ICM. The launch of satellites such as \textit{XMM-Newton} and \textit{Chandra}, with improved spatial and spectral resolutions, enabled more precise measurements of elemental abundances, and allowed determination of supernovae contribution to the metal enrichment in galaxy clusters' cores out to R$_{500}$\setcounter{footnote}{0}\footnote{R$_{500}$ is the radius at which the mean density of the cluster is 500 times the critical density of the Universe at the cluster's redshift.}\citep{buote03,werner06,deplaa07,matsushita07b,baldi07}. The subsequently discovered centrally peaked Fe abundance at the center of cool core clusters may be explained by a more extensive period of enrichment by SN Ia explosions in the brightest cluster galaxy \citep{degrandi2004,bohringer2004}. This Fe enhancement in cluster cores is also seen in high spatial resolution observations with {\it XMM-Newton} \citep{simi2009, b12b,degrandi14}.

Studies of azimuthal spatial distributions of metal abundances out to cluster outskirts have become possible with the launch of {\it Suzaku}. Due to its low particle background, deep observations of clusters of galaxies with \textit{Suzaku} provide the measurements of elemental abundances and SN ratio out to R$_{200}$ in nearby clusters ($z<0.02$), e.g., the Perseus and Virgo clusters \citep{werner13,simi15}. These results suggest a uniform distribution of SN Ia and SN cc yields in the cluster outskirts, thus favoring an early enrichment by SN Ia started in the early stages of the cluster formation.

\begin{table*}[ht!]
\centering
\caption{\textit{Suzaku} and \textit{Chandra} observations of Abell 3112.}
\scalebox{0.9}{
{\renewcommand{\arraystretch}{1.0}
{ \small \begin{tabular}{ccccccccc}
\hline \hline\\
Instrument&Obs. ID &  Pointing    &   RA    &   Dec  &  Observation & Exp. & Filtered Exp. & PI\\
                  &             &                   &           &            &                      &XIS0/XIS1/XIS3& XIS0/XIS1/XIS3 &  \\
   & $~$     & $~$     &    (J2000)  &   (J2000)    &Date&  (ks)    & (ks)   \\ \hline\hline
   \\
\textit{Suzaku} XIS&803054010 & On-axis & 49.478 & -44.248 & 2008 May 23 &67.5/67.5/67.5 & 54.9/54.9/54.9 & M. Bonamente\\
                    &808068010 & On-axis & 49.498 & -44.251& 2013 Jun 23 &119.1/119.1/119.1 & 113.5/113.5/113.5 & E. Bulbul \\
                    &808068020 & On-axis & 49.497 & -44.236 & 2013 Jun 25 & 65.4/65.4/65.4 & 54.4/54.4/54.4 & E. Bulbul  \\
                    &809116010& Offset & 49.354 & -44.489 & 2014 Dec 09 &107.9/107.9/107.9 & 87.3/87.3/87.3 & E. Bulbul \\
                    &809116020& Offset & 49.354 & -44.449 & 2014 Dec 12 & 97.9/97.9/97.9 & 80.4/80.4/80.4 & E. Bulbul \\
                    \\
\textit{Chandra} ACIS-I & 13135 & On-axis& 49.481 & -44.258 & 2011 Mar 14 & 42.8 & 42.2 & S. Murray\\
			  & 6972 & Offset & 49.421 & -44.410 & 2006 Apr 18 & 30.2 & 29.7 & M. Markevitch\\
			  \\
			  
\hline
\end{tabular}}}}
\label{tab:obs}
\end{table*}

Extending these studies to more distant clusters has become possible through deep \textit{Suzaku} observing programs. Abell 3112 (hereafter Abell 3112) is one such object, an archetypal cool-core cluster at redshift 0.075. The cluster has a strong radio source, PKS 0316--44, located in the cluster center \citep{abell1}. The mass deposition rate of $10^{+7}_{-5}\ M_{\odot}\ yr^{-1}$ indicated by {\it XMM-Newton} observations is much less than the expected rate from cooling flow clusters \citep{odea08,b12a}. It was also reported that a soft X-ray gas was present in the ICM above the contribution from the diffuse 4--5 keV hot gas. This soft excess was first thought to be well described with an additional non-thermal power-law model or with a 1~keV thermal model of low metal abundance \citep{nevalainen03, bonamente07, lehto10}. However, \citet{b12a} ruled out the thermal origin of this soft excess using {\it XMM-Newton RGS} observations, leaving the possibility for non-thermal interpretation of a potential population of relativistic electrons with $\sim$7\% of the cluster's gas pressure. The peaked  Fe, Si, and S abundances in the core region reported in \citet{b12a,b12b} imply an ongoing SN Ia contribution towards the immediate cluster core ($<0.5^{\prime}$) followed by a more uniform SN cc contribution. Finally, \citet{b12b} used higher resolution {\it XMM-Newton RGS} observations of Abell 3112 to constrain the SNe models using a new method, \textit{snapec}, and reported that 30.3$\% \pm 5.4\%$ of the total SN which enriched the ICM are SN Ia within the immediate core ($\sim50$ kpc) of the cluster. It was also reported that the total number of SN explosions required to create the observed metals is (1.06 $\pm$ 0.34) $\times10^{9}$ in the cluster core \citep{b12b}.

In this paper, we take a further step to investigate the radial distribution of SN enrichment in Abell 3112 out to the cluster's virial radius by comparing deep {\it Suzaku} and {\it Chandra} X-ray observations with the nucleosynthesis models available in the literature. The paper is organized as follows: we describe {\it Suzaku} and {\it Chandra} data analysis in Section \ref{sec:reduction}. In Section \ref{sec:modeling}, we give an overview of spectral extraction and background modeling. The systematic uncertainties relevant to {\it Suzaku} analysis are described in Section \ref{sec:syst}. We provide our results and conclusions in Sections \ref{sec:results} and \ref{sec:conc}. 

At the cluster's redshift, $1^{\prime}$ corresponds to  $\sim$ 82 kpc. The cosmological parameters used in the analysis are H$_{0}= 73$  km s$^{-1}$ Mpc$^{-1}$, $\Omega_{\text{M}} = 0.27$, $\Omega_{\Lambda} =  0.73$. Unless otherwise stated, reported errors correspond to 68\% confidence intervals.

\section{Observations and Data Reduction}
\label{sec:reduction}
\subsection{\textit{Suzaku} Data Reduction}
\begin{figure*}
\centering
\includegraphics[width=0.93\textwidth]{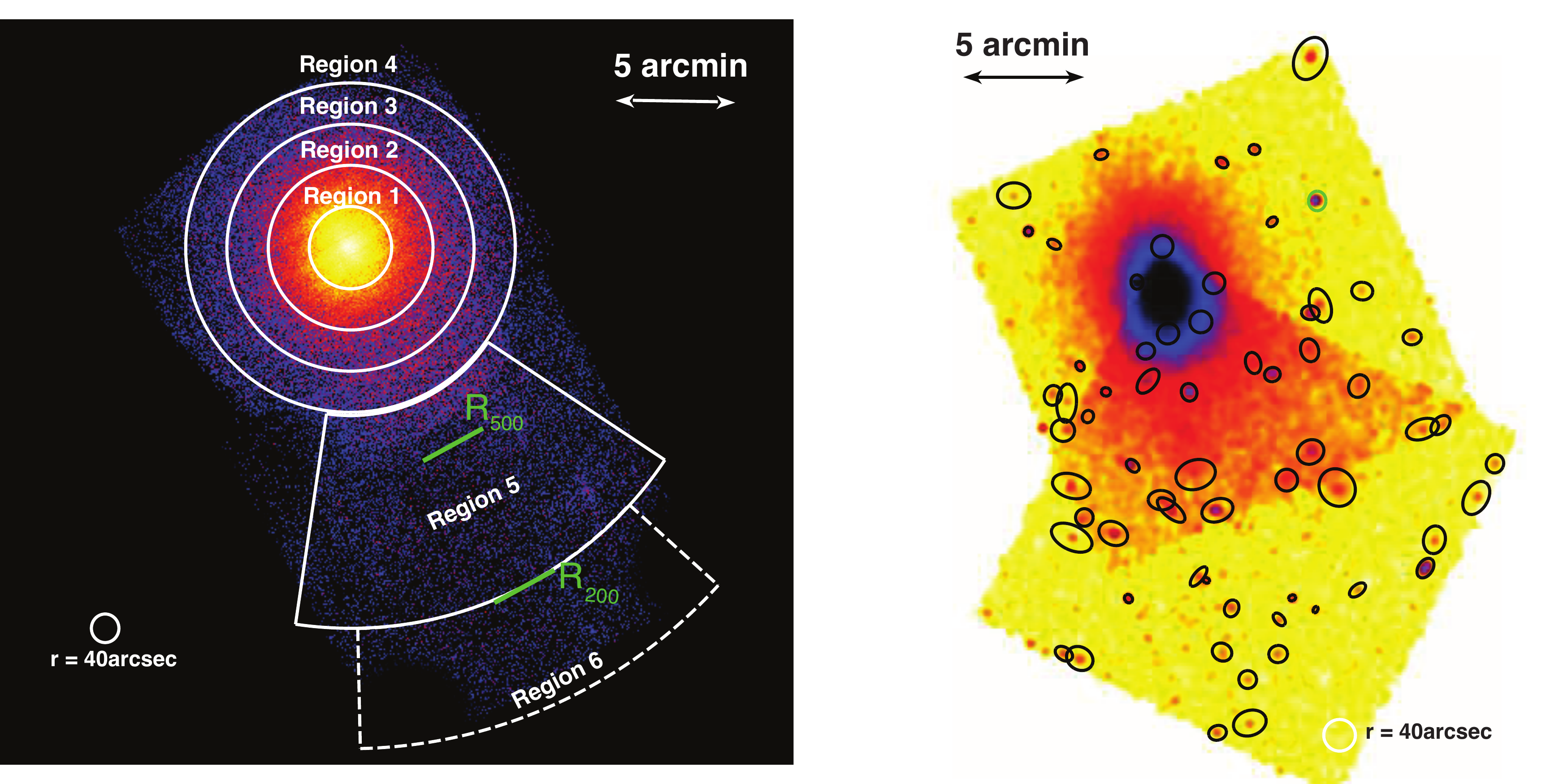}
\caption{{\bf Left panel:} Exposure corrected, NXB background subtracted {\it Suzaku} XIS image of Abell 3112. The image is extracted in the 0.5 $-$ 7 keV energy range. The spectral extraction regions out to R$_{200}$  are shown in white. The region which is used to extract the local background spectrum is shown in dashed lines. The over density radii R$_{500}$ and R$_{200}$ are marked with green bars in the figure. The point source exclusion radius of 40$^{\prime\prime}$ is shown in the lower corner of the image. {\bf Right panel:} Background subtracted \textit{Chandra} image of Abell 3112 is given in 0.5 $-$ 7.0 keV energy band. \textit{Chandra} pointings are used to detect point sources within the {\it Suzaku} FOV.  The brightest point source, which is used in estimating point-source exclusion extent, is shown in the green.}
 \vspace{1mm}
\label{fig:img}
\end{figure*}

Abell 3112 was observed with \textit{Suzaku} with five pointings between May 2008 and Dec 2014.
The unfiltered \textit{Suzaku} data are analyzed by using HEASOFT version 6.17 and the latest calibration database (CALDB)  as of  November 2015. Here, we summarize the data analysis steps briefly. The details of {\it Suzaku} data reduction are described in \citet{bulbul16a, bulbul16b}. The FTOOL \textit{aepipeline} is used to reprocess the unfiltered event data files using the latest calibration and screening criteria. Additionally, we require elevation angles above 5$^{\circ}$ and 20$^{\circ}$ for the night and day Earth rim and the geomagnetic cut off rigidity of $>6$~GV. The data taken when the satellite passes through the regions affected by South Atlantic Anomaly (SAA) and the  $^{55}$Fe calibration sources at two far corners of CCD chips are excluded from the analysis. The event files in the 3 $\times$ 3 and 5 $\times$ 5 editing modes are combined. An additional correction for the comparable fraction of flickering pixels is applied to the data which are taken after Jan 2014\footnote{\url{http://www.astro.isas.jaxa.jp/suzaku/analysis/xis/nxb\_new/}}. The filtered exposure times are given in Table \ref{tab:obs}. A total of 1.2~Ms total filtered \textit{Suzaku} XIS exposure time (391~ks exposure per XIS detector) is used in this analysis. The non-X-ray background (NXB) images are generated using the `night-Earth' data (NTE) using the FTOOL \textit{xisnxbgen} \citep{tawa2008}. The NXB images are then subtracted from the mosaicked image prior to exposure correction. We generate the exposure maps as described in \citet{bautz2009} and \citet{bulbul16a, bulbul16b} using {\it xissim} and {\it xisexpmapgen}. Before exposure correction is applied, underexposed regions with $< 15$\% of the maximum exposure time are removed. An exposure corrected and particle background subtracted {\it Suzaku} mosaic image is shown in the left panel of Figure \ref{fig:img}.

\subsection{Chandra Data Reduction}
\label{sec:chandra}

To detect X-ray point sources unresolved by {\it Suzaku}, we use the two overlapping {\it Chandra} pointings of the cluster.
\textit{Chandra} ACIS-I data are filtered from background flares using LC$\_$CLEAN through \textit{Chandra} analysis software CIAO version 4.7 with CALDB version 4.6.7. The filtered light curves show no left-over significant background flares. The filtered exposure times are given in Table \ref{tab:obs}. We extract image in the 0.5 $-$ 7 keV band. The background image is extracted from the blank-sky observations. To account for variations in the particle background, we use count rates detected in the 9 $-$ 12 keV band to match Abell 3112 observations as described in \citet{markevitch03}. The CIAO's {\it wavdetect} tool is used to determine the locations of the point sources in the field of view (FOV). The point sources detected by {\it Chandra} in the {\it Suzaku} FOV are shown in the right panel of Figure \ref{fig:img}.

\section{Spectral Modeling and Background Subtraction}
\label{sec:modeling}
Eliminating the contribution from local foreground, extra-galactic background, and the point sources within the {\it Suzaku} FOV is crucial in studies of cluster outskirts. In this section, we first describe our spectral fitting procedure for cluster emission. Additionally, we  describe our background subtraction and point source optimizations methods in Subsections \ref{sec:ps} and  \ref{sec:bkg}.

\subsection{Cluster Emission Modelling}

To examine the spectral properties of Abell 3112, we extract spectra in five regions  surrounding the cluster's centroid from the filtered event files in \textit{XSELECT} for each XIS sensor (see Figure \ref{fig:img}). The selected regions cover a radial range from cluster core out to the virial radius ($R_{200}$) (Region 1, 0$^{\prime}$-2$^{\prime}$;  Region 2, 2$^{\prime}$-4$^{\prime}$; Region 3, 4$^{\prime}$-6$^{\prime}$; Region 4, 6$^{\prime}$-8$^{\prime}$; and Region 5, 8$^{\prime}$-18$^{\prime}$). The regions are selected based on the total source counts ($>10^{4}$ counts) in each. The over density radii R$_{500}$ and R$_{200}$ are also marked in Figure \ref{fig:img}. The FTOOLS \textit{xissimarfgen} and \textit{xisrmfgen} are used to generate the effective area ancillary response file (ARF) and detector redistribution matrix file (RMF), respectively. For each annulus and each observation, we merge data from front-illuminated (FI) XIS0 and XIS3 detectors. The back-illuminated (BI) XIS1 data are fit simultaneously with the FI spectra. Spectral fitting is performed in the 0.7 $-$ 7 keV energy band where the {\it Suzaku} XIS detectors are the most sensitive. The cluster emission is modeled with ATOMDB version 2.0.2 \citep{smith2001, foster2012}. \textit{XSPEC} v12.9.0 is used to perform the spectral fits with the extended C-statistic as an estimator of the goodness-of-fits \citep{arnaud1996}.

The soft local foreground and cosmic X-ray background parameters are fixed to the best-fit values obtained from the joint fit of the ROSAT All Sky Survey ({\it RASS}) and local background as described in Section \ref{sec:bkg}. The particle background spectra are subtracted prior to fitting. The spectra are fit with a single-temperature thermal model (1T {\it apec} or {\it snapec}) with free temperature, metallicity, and normalization. We also search for two-temperature structure by adding a second thermal component  (2T {\it apec}). The Galactic Column density is fixed to the LAB value of 1.33$\times 10^{20}$ cm$^{-2}$ in our fits \citep{kalberla2005} and solar abundances adopted from \citet{solar}. The redshift is fixed to 0.075 \citep{braglia11}.

Cutoff-rigidity-weighted non X-ray background (NXB) spectra are extracted from the night-time-earth data for each detector using the \textit{xisnxbgen} tool. NXB event files are reprocessed following the same procedure described in Section \ref{sec:reduction}. The same annular sections are used to produce NXB spectra in {\it XSELECT} after the calibration sources are removed.

 \subsection{Point Sources Optimization}\label{sec:ps}
The main obstacle in excising point sources in analyses of {\it Suzaku} observations is the relatively large size of point-spread-function (PSF) of the {\it Suzaku} mirrors.
We use the two overlapping {\it Chandra} observations (both on-axis and offset) to detect X-ray point sources unresolved by \textit{Suzaku} (see Section \ref{sec:chandra}). The PSF sizes of {\it Suzaku} and {\it Chandra} are quite different. Therefore, the extents of the point sources detected by {\it wavdetect} using {\it Chandra} observations cannot be used directly to exclude point sources in the \textit{Suzaku} FOV. We use the same procedure described in detail in \citet{bulbul16a} to determine a conservative exclusion radii for point sources detected by {\it Chandra} pointings. We selected the brightest point source in the {\it Suzaku} FOV (J2000; R.A.: 49.342$^\circ$; DEC: -44.173$^\circ$), which is located in a fairly faint region of the cluster (shown in the green circle in Figure \ref{fig:img}). The \textit{Chandra} spectrum of the point source is extracted using the {\it specextract} tool in {\it CIAO}. The spectrum of the source is fitted with an absorbed power-law with a fixed index set to 1.4 and variable normalization \citep{hickox2006}. We then simulate {\it Suzaku} observations of the point source based on the best-fit flux ($3.47 \times 10^{-5}$  photons keV$^{-1}$ cm$^{-2}$ s$^{-1}$) and on the power-law index (1.4) obtained from the {\it Chandra} observations using the FTOOL {\it xissim}. To estimate the effect of the point source contamination to the surrounding cluster ICM gas, we add  simulated diffuse emission to the spectrum with a total net counts of 2000.  Our goal is to measure the plasma temperature with better than $<$20\% accuracy in these simulations. We extract the spectrum around the point source with incremental extraction radii to determine the radius where the cluster emission is not affected by the point source contamination. We find that excluding $r>40^{\prime\prime}$ around the point source has a minimal effect on the cluster plasma temperature, metallicity, and normalization. Since this point source is in a faint region of the cluster and all our spectra include at least 10$^{4}$ counts, the exclusion radius of $40^{\prime\prime}$ is a conservative estimate for all point sources detected by \textit{Chandra} observations in the {\it Suzaku} FOV.  The exclusion radius is shown in the lower left corner of Figure \ref{fig:img} (left panel).

\subsection{Modeling of the Local X-ray Background} \label{sec:bkg}
Understanding temporal and spatial variations in the local X-ray background is crucial in analyses of faint cluster outskirts.
The variable soft X-ray background must be examined carefully before the spectral fits are performed. We first extract a local background spectrum from the outermost region (Region 6 in Figure \ref{fig:img}, 18$^{\prime}-24^{\prime}$ which is beyond R$_{200}$) where the expected contribution from the cluster thermal emission is minimal. We also extract the {\it RASS} data from a 1$-$2 degree annulus surrounding the central sub-cluster's centroid\footnote{\url{http://heasarc.gsfc.nasa.gov/cgi-bin/Tools/xraybg/xraybg.pl}}. The {\it RASS} background spectrum is simultaneously fit with the local background XIS FI and BI spectra. The local X-ray background model consists of two absorbed thermal components ({\it apec}) for the Galactic Halo (GH; E$\sim$0.25~keV) and the Hot Foreground (HF; E$\sim$0.75~keV), an unabsorbed thermal model for the Local Hot Bubble (LHB; E$\sim$0.1~keV), and a power-law component for unresolved point sources (cosmic X-ray background; CXB) with a photon index of 1.4 \citep{hickox2006}. We note that we use the full energy band between 0.5 $-$ 7 keV band for XIS spectra and 0.5 $-$ 2 keV for RASS spectra in background fits in order to have a better handle on the soft GH component. In order to avoid degeneracies between the three thermal background components, we fix the temperatures to the values reported in \citet{Snowden2008} and \citet{b12a}. We also add two Gaussian models to eliminate the O {\sc VII} and O {\sc VIII} lines from solar wind charge exchange at 0.56 keV and 0.65 keV. The metallicities of these {\it apec} models are set to solar, while the redshifts are fixed at zero. We find a good-fit with $C-$stat value of 507.3 for 341 d.o.f. The best-fit values of the background model are given in Table \ref{tab:local}. The best-fit normalization of the power-law is 1.41$_{-0.14}^{+0.14}\times 10^{-7}$~photons~keV$^{-1}$~cm$^{-2}$~s$^{-1}$~arcmin$^{-2}$ at 1 keV, corresponding to a CXB flux of 4.3 $+/-$ 0.4 $\times 10^{-12}$ ergs s$^{-1}$ cm$^{-2}$ deg$^{-2}$ in the 0.5--2keV band.

\begin{threeparttable}[h!]
\centering
\scriptsize
\caption{Best-fit Parameters from Fits to Soft X-ray Background}
{\renewcommand{\arraystretch}{1.5}
\begin{tabular}{llcccl}
\hline \hline
 Component 	& \textit{kT}	& Normalization    & Flux (0.5 $-$ 2.0 keV)\\	
 			&  (keV)  		& $10^{-8}$ cm$^{-5}$ & $10^{-16}$ ergs s$^{-1}$ cm$^{-2}$ \\
 \hline
 GH		&  0.25$^{*}$  		&  0.80$_{-0.60}^{+0.70}$& 0.14$_{-0.09}^{+0.09}$\\
 HF 		&  0.75$^{*}$		& 1.20$_{-0.30}^{+0.30}$&0.34$_{-0.07}^{+0.07}$\\ 
 LHB		& 0.10$^{*}$		&  68.4$_{-6.40}^{+6.20}$& 1.29$_{-0.11}^{+0.12}$\\
 \hline
\end{tabular}}
\begin{tablenotes}
      \small
      \item Note: $^{*}$ indicates fixed parameters in the background fits. The temperatures are fixed to value reported in \citet{Snowden2008}. 
      \end{tablenotes}
\label{tab:local}
\end{threeparttable}
\section{Systematic Uncertainties}
\label{sec:syst}

The analyses of low-surface brightness regions of clusters with {\it Suzaku} may be subject to systematic uncertainties. To estimate the magnitude of these, we consider the following potential sources of uncertainties: i) systematics associated with the CXB level; ii) systematics due to variations in the soft X-ray and particle background; iii) contamination due to stray light and the large size of the PSF of {\it Suzaku}'s mirrors. We describe how we estimate and handle these in detail below.

\begin{figure*}[]
\centering
\includegraphics[width=1.0\textwidth]{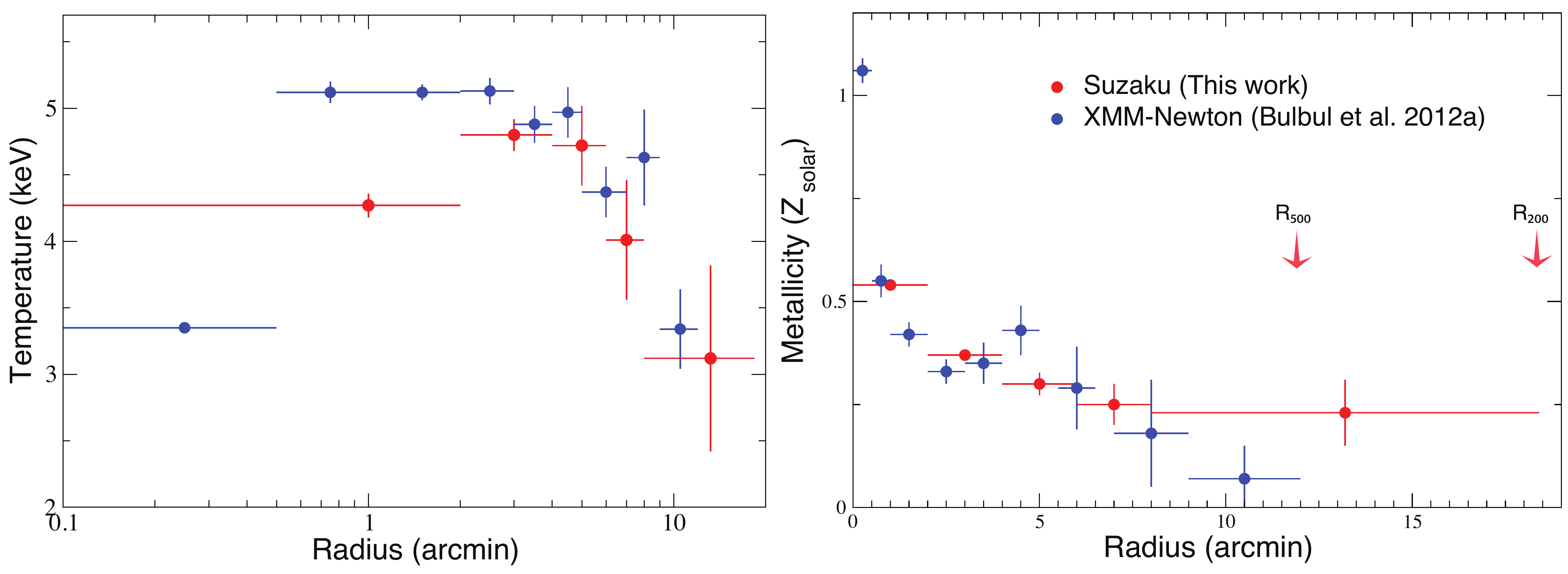}
\caption{ The radial profile of temperature (left panel) and  metallicity (right panel) for Abell 3112 obtained from a single temperature \textit{apec} model. The statistical errors (68\% confidence level) together with systematics on the {\it Suzaku} results are over plotted. {\it Suzaku} results are compared with the {\it XMM-Newton} results reported by \citet{b12a}. The {\it XMM-Newton} and {\it Suzaku} measurements are in agreement with each other at 1$\sigma$ level. While {\it XMM-Newton} observations are able to determine the temperature and metallicity from the core out to R$_{500}$, we are able to measure these parameters out to $R_{200}$ of the cluster.
 }
\label{fig:profiles}
\end{figure*}
\subsection{The Cosmic X-ray Background}
The variations in the unresolved CXB within the XIS FOV can be a source of serious systematic uncertainty. Following the same approach in \citet{bulbul16a}, we find that the detection limit in our observations is $6.7 \times 10^{-14}$ ergs cm$^{-2}$ s$^{-1}$ in the 2.0 $-$ 10.0 keV energy band. The contribution of unresolved point sources to the total flux is calculated using the formula given in \citet{moretti03}:

\begin{equation}
\begin{aligned}
F_{CXB}=  2.18\pm0.13 \times 10^{-11}- &\int_{S_{excl}}^{S_{max}} (\frac{dN}{dS}) \times S~dS \\
& \rm {ergs \  cm^{2} \ s^{-1} \ deg^{-2}}
\end{aligned}
\label{eqn:cxb1}
\end{equation}

\noindent where the cumulative number of point sources per flux indicated as dN/dS is integrated 
over the detection limit from the lower bound $S_{excl}=6.7 \times 10^{-14}$ ergs cm$^{-2}$ s$^{-1}$ to the upper bound $S_{max}=8.0\times10^{-12}$ ergs cm$^{-2}$ s$^{-1}$ given in \citet{moretti03}. We use a total flux of $2.18\pm0.13 \times 10^{-11} ~ \text{erg cm}^{-2} ~ \text{s}^{-1} ~ \text{deg}^{-2 }$ obtained from \textit{Swift} data \citep{moretti09}. This flux is also consistent with the total CXB flux from {\it Chandra} and {\it XMM-Newton} observations \citep{moretti03, deluca04}. For the populations of the point sources we adopt an analytical model provided in \citet{moretti03}:

\begin{equation}
N(>S)= N_0\Big[\frac{(2\times10^{-15})^\alpha}{S^\alpha+S_0^{\alpha-\beta}S^\beta}\Big].
\label{eqn:cxb2}
\end{equation} 

\noindent where the best-fit parameters are $N_{0}=5300^{+2850}_{-1400}$, $S_{0}=(4.5^{+3.7}_{-1.7}) \times 10^{-15}$ ergs cm$^{-2}$ s$^{-1}$, $\alpha=1.57^{+0.10}_{-0.08}$, and $\beta=0.44^{+0.12}_{-0.13}$. Using the best-fit parameters of hard energy band given in \citet{moretti03}, we find that the unresolved flux contribution in the 2 $-$ 10 keV band to the  CXB flux is $1.38 \pm 0.62 \times 10^{-11}$  ergs cm$^{-2}$ s$^{-1}$ deg$^{-2}$. 

\begin{table}[h!]
\centering
\caption{Estimated $1\sigma$ Fluctuations in the CXB level due to unresolved point sources in the \textit{Suzaku} FOV in units of 10$^{-12}$ ergs cm$^{-2}$ s$^{-1}$ deg$^{-2}$.}
{\renewcommand{\arraystretch}{1.5}
\scriptsize
{\begin{tabular}{cccccc}
\hline\hline
		&  Region 1 & Region 2 & Region 3 &  Region 4 & Region 5 \\ \hline 
CXB & 10.10 & 5.78 & 4.48 & 3.78 & 2.12\\
Fluc. \\
\hline
\end{tabular}}}
\label{table:cxb}
\vspace{2mm}
\end{table}

The deviations from the expected CXB level due to the unresolved point sources are,
\begin{equation}
{\sigma_B}^2=\frac{1}{\Omega}\int_0^{S_{excl}} (\frac{dN}{dS}) \times S^2~dS,
\label{eqn:cxb3}
\end{equation} 

\noindent where $\Omega$ is the solid angle. We then calculate $1\sigma$ RMS CXB fluctuations using Equations \ref{eqn:cxb1} and \ref{eqn:cxb2} for each region. The results are shown in Table \ref{table:cxb}. We find that a typical 1$\sigma$ uncertainty on the measured CXB level is comparable to the RMS value of CXB fluctuations. We note that this uncertainty is used in Section \ref{subsec:montecarlo} to account for the CXB variations. These systematics are included in the final systematic errors on the observable quantities.

\subsection{Systematics due to Variations in the Soft X-ray and Particle Background}
\label{subsec:montecarlo}

We model the soft X-ray background by jointly fitting the ROSAT data with the local X-ray background spectra (including LHB, GH, HF), and CXB obtained from the annuli encompassing the 18$^\prime$-24$^\prime$ region. To take into account the spatial variations which may dominate the background, we perform 1000 Markov chain Monte Carlo (MCMC) realizations of the best-fit background model. The model parameters are allowed to vary within their $1\sigma$ uncertainty range. An uncertainty up to 3.6\% on the NXB level is also taken into account in these realizations \citep{tawa2008}. We find that the systematic variations in the soft foreground, CXB, and the particle background level has an effect of $<$ 1\% on the best-fit temperatures and normalizations of Regions 1 and 2. Variations up to 2\%, 6\%, and 16\% are measured in Regions 3, 4, and 5. The uncertainties due to the variation in the soft X-ray background are taken into account in the total error budget calculations by adding them in quadrature.

\begin{table}[h!]
\centering
\caption{Percentage Contribution of PSF Scattering}
\begin{threeparttable}
{
\scriptsize
{\renewcommand{\arraystretch}{1.5}
{\begin{tabular}{cccccc}
\hline\hline
		&  Region 1 & Region 2 & Region 3 &  Region 4 & Region 5 \\ \hline 
Region 1  & 68.5 &15.6 & 3.47 &1.33 & 0.33 \\
Region 2  & 14.3 & 65.6  & 16.4 & 2.25  & 0.38  \\
Region 3  & 0.24 &17.7 &  64.5  & 13.7 & 0.82\\
Region 4  & 0.26 &1.52 & 15.2 &  66.8 & 6.81 \\
Region 5  & 0.09 & 0.27 & 0.79 & 7.06 & 89.2\\
\hline\hline
\end{tabular}}}}
\end{threeparttable}
\label{table:psf}
\end{table}

\subsection{Systematics due to Scattered light and PSF Scattering}

The relatively large size of the {\it Suzaku} mirror PSF ($\sim2^{\prime}$) may cause photons emitted from a particular region in the sky to be detected elsewhere on the detector. We note that the size of each spectral extraction region used in this work is larger than the PSF size, minimizing the effect of PSF scattering in this analysis. To estimate the magnitude of this uncertainty, we use the ray-tracing simulator {\it xissim} to generate {\it Suzaku} event files using \citet{ishisaki07}. The {\it Chandra} ACIS images and the best-fit {\it Suzaku} spectral models are used to simulate event files of each XIS sensor with $1\times 10^{6}$ photons. The image of each sector shown in Figure \ref{fig:img} (left panel) is extracted from the simulated event files. The percentile contribution of the flux on each sector from adjacent regions are shown in Table \ref{table:psf}. Columns refer to the percentage fluxes providing the flux, while the rows refer to the percentage fluxes receiving the flux in Table \ref{table:psf}. For instance, 3.47\% (first row, third column) is the fraction of photons leaked from Region 1 contaminating Region 3. As clearly seen in Table \ref{table:psf}, most of the photons which originate from one particular annulus in the sky are detected in the same region on the detector, while up to 17\% of the photons may be detected in surrounding annuli. However, the fraction of photons which are detected in the outermost annulus that scatter from the inner regions is negligibly small ($<1\%$). The results are consistent with the fractions reported in \citet{bautz2009} and \citet{bulbul16a}.

To estimate the effect of the PSF scattering and the scattered light contribution to the variables, e.g. temperature and metallicity, we jointly fit the spectra of each sector with the normalizations scaled according to the reported fractions in Table \ref{table:psf}. Although the uncertainty on the measured temperature is smaller than the statistical errors in each sector, we added these in quadrature to the total error budget of the thermal model variables.

\begin{figure}[]
\centering
\includegraphics[width=0.48\textwidth]{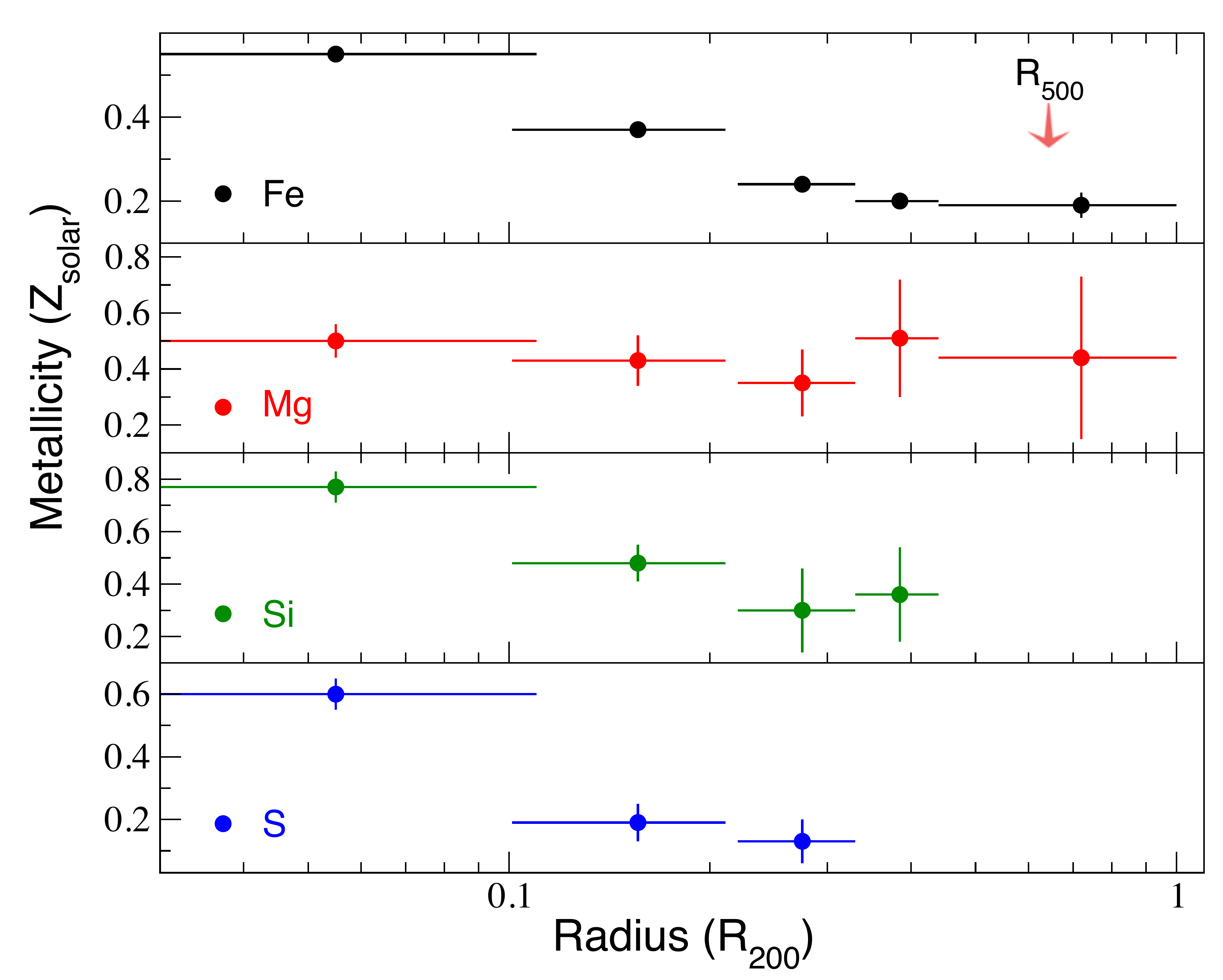}
\caption{The radial distribution of elemental abundances of $\alpha$-elements, Si, S, Mg, and Fe. We are able to determine Mg and Fe abundances out to R$_{200}$. The uncertainties are for $\Delta C$=1. }
\label{fig:elementprofiles}
\vspace{3mm}
\end{figure}
\section{Results}
\label{sec:results}

We start with examining the global  properties of the cluster by modeling the five spectra using the thermal models as described in Section \ref{sec:modeling}. The results for the chemical enrichment are also described in this Section.

\subsection{Global Spectral Properties}

To examine the global temperature and metallicity out to $R_{200}$ of Abell 3112, we first fit the spectra with a 1T {\it apec} model. The model parameters between different observations are tied to each other. The best-fit projected temperature  together with their systematic and statistical uncertainties are shown in Figure \ref{fig:profiles} and Table \ref{table:temp}. In the same figure, the {\it Suzaku} results are compared with the previous measurements from the {\it XMM-Newton} observations \citep{b12a}. The {\it Suzaku} and {\it XMM-Newton} results are in agreement with each other at $1\sigma$ confidence level from the cluster core out to R$_{500}$. The plasma temperature in the core (4.27~$\pm$~0.01 keV) is cooler than the temperature at intermediate radii, confirming that Abell 3112 is a cool core cluster. While previous observations were able to measure the temperature only out to R$_{500}$, we are able to measure the ICM temperature to R$_{200}$ owing to {\it Suzaku}'s lower particle background.

\begin{table}[htbp]
\caption{The best-fit Parameters of the 1T {\it APEC} Model\label{table:temp}}
\label{Tab:ApecResults}
\begin{center}
\renewcommand{\arraystretch}{1.5}
\scriptsize
\begin{tabular}{cccccc}
\hline
Region & \textit{kT} & Metallicity & $N$ &\textit{C}-stat\\

&(keV) & ($Z_{\odot}$)	& (10$^{-6} $cm$^{-5}$) & (dof)\\ \hline

Region 1 &4.27$\pm$ 0.02  & 0.54 $\pm$ 0.01 & 881.0 $\pm$ 2.4 & 1260.9 (849) \\
Region 2 &4.81$\pm$ 0.04& 0.36 $\pm$ 0.01 &  204.1 $\pm$ 2.9 & 1383.6 (1135)  \\ 
Region 3 &4.66 $\pm$ 0.05 & 0.26 $\pm$ 0.03 & 71.7 $\pm$ 0.14 & 997.3 (844) \\ 
Region 4  &4.26 $\pm$ 0.10 & 0.25 $\pm$ 0.05 & 22.3 $\pm$ 2.3 & 397.9 (279)\\
Region 5 & 3.37 $\pm$ 0.77 &  0.22 $\pm$ 0.08 & 1.1 $\pm$ 0.2  & 264.9 (136)\\
\hline
\end{tabular}
\end{center}
\end{table}

In order to investigate the multi-phase gas in the ICM, we fit the spectra with a 2T {\it apec} model. The best-fit model parameters are given in Table \ref{table:temp2t}. The best-fit temperature of the ICM in Region 1 becomes 5.86 $\pm$ 0.37 keV, with a lower \textit{kT} component of 3.23$_{-0.14}^{+0.07}$ keV. The temperature in Region 2 is $5.63^{+0.60}_{-0.26}$ with a lower \textit{kT} component of $3.41^{+0.42}_{-0.21}$ keV. The metallicity remains unchanged in both regions with an addition of the second {\it apec} model.  Adding the second thermal component decreases $\Delta$C-stat of the fit to the spectra of Region 1 and Region 2  ($\Delta$C-stat=126 for 2 dof in Region 1; $\Delta$C-stat=17 for 2 dof in Region 2). C-statistics does not provide a direct statistical test to quantify the significance of the improvement in adding the secondary {\it apec} component. We therefore calculate the corresponding $\chi^{2}$ values from the best-fits (which are obtained using C-statistics). The improvement in $\chi^{2}$ values are 114 and 12 in Region 1 and Region 2 for two extra dof (temperature and normalization of the second {\it apec} model). This corresponds to F-test values of 40.8 and 4.9 with null hypothesis probabilities of $10^{-25}$ and 0.7\% in Regions 1 and 2. Comparing the normalizations of the 2T {\it apec} models, both components are equally contributing the total emission (see Table \ref{table:temp2t}). The temperatures measured in 0--2$^\prime$ region of {\it Suzaku} observations are consistent with the temperatures reported from {\it XMM-Newton} observations in the cluster core \citep{b12a}. The limited statistics of the spectra extracted from Regions 3, 4, and 5 do not allow to test the multi-phase nature of the plasma in these regions. We, therefore, do not provide the 2T results from those fits here.

\begin{table}[htbp]
\caption{The best-fit Parameters of the 2T {\it APEC} Model\label{table:temp2t}}
\label{Tab:ApecResults}
\begin{center}
\renewcommand{\arraystretch}{1.5}
\normalsize
\begin{tabular}{lcc}
\hline
Parameter	 &	Region 1 & Region 2 \\
\hline
$kT_{1}$ (keV) &  3.24 $_{-0.16}^{+0.20}$  & 3.41$_{-0.09}^{+0.07}$\\
$N_{1}$ ($10^{-4}$ cm$^{-5}$) & 4.32$_{-0.78}^{+0.90}$ & 0.65$_{-0.06}^{+0.08}$\\
$kT_{2}$ (keV) &  5.94 $_{-0.15}^{-0.16}$ & 5.63 $\pm$ 0.10  \\
$N_{2}$ ($10^{-4}$ cm$^{-5}$)  & 4.46$_{-0.77}^{+0.78}$ &  1.41$_{-0.26}^{+0.15}$ \\
Metallicity ($Z_{\odot}$) & 0.55 $\pm$ 0.01 & 0.37 $\pm$ 0.01  \\
$C$-stat& 1135.52 (847) & 1366.25 (1133) \\
\hline
\end{tabular}
\end{center}
\end{table}

We also compare metallicity profiles obtained from {\it XMM-Newton} and {\it Suzaku} observations in Figure \ref{fig:profiles}. While {\it XMM-Newton} observations can accurately constrain the profiles in the core of the cluster out to $\sim$ 0.5R$_{500}$, {\it Suzaku} observations are able to constrain metallicity at radii out to $R_{200}$. The regions which are covered by both {\it XMM-Newton} and {\it Suzaku} observations are in agreement with each other at the $1\sigma$ level. 

The metallicity profile is peaked at the center, and remains fairly constant beyond $\sim$0.5R$_{500}$. The overall metallicity of the ICM (mostly driven by the Fe lines) in the cluster outskirts is 0.25 $\pm$ 0.05  Z$_{\odot}$ and 0.22 $\pm$ 0.08 Z$_{\odot}$ in Regions 4 and 5. which cover the region from 0.5R$_{500}$ to R$_{200}$. These values are consistent with metallicities` measured in the outskirts of low-mass clusters \citep{bulbul16a,fujita2008}. We further investigate the radial abundance distributions of individual $\alpha$-elements, such as Si, S, Fe, and magnesium (Mg) out to $R_{200}$ (see Figure \ref{fig:elementprofiles}). The fits are performed with a single temperature {\it vapec} model. The Fe, Si, S, and Mg elemental abundances are allowed to vary independently while other elemental abundances which cannot be measured, (e.g. carbon and argon) are fixed to the measured Fe abundance at the outskirts, 0.25$Z_{\odot}$. We find that Si, S, and Fe abundances show an increasing trend towards the core of the cluster, confirming the results from {\it XMM-Newton} observations. We are also able to extend the detection of Si out to $\sim0.5R_{200}$ in the {\it Suzaku} observations, while previous {\it XMM-Newton} observations report the detection of these  metals only in the very central region ($<0.06R_{200}$) \cite[see][]{b12a}.

SN Ia produce large amounts of Fe, while lighter elements, such as Mg, is produced mainly by SN cc. Measurements of the radial profile of Mg and Fe abundances within the ICM provide clues on the relative contribution of different types of supernovae to the chemical enrichment. For instance, SN cc products are thought to be produced in early in the formation history of clusters at z$\sim$2--3 \citep{simi15}. We, therefore, investigate distributions of these elements in the outskirts of Abell 3112. We perform fits  to the observed Mg and Fe profiles with phenomenological models to quantify the change in their distribution with radius. In a power-law model fit to the observed Mg profile, the best-fit normalization is 0.35$\pm$0.12 and index of 0.11$\pm$0.15. We find a good fit with overall $\chi^{2}$ of 0.62 (3 d.o.f.). Although, the power-law index in this fit indicate a slight decline with radius, it is consistent with zero. This indicates that the distribution of the Mg is consistent with a uniform profile. However, due to large uncertainties  neither the uniform profile, nor the peaked profile can be excluded based on the {\it Suzaku} data.

We observe a steeper decline in the Fe abundance profile with power-law index of 0.54$\pm$0.15, normalization of 0.13$\pm$0.03. The overall $\chi^{2}$ is 1.37  for 3 d.o.f. The observed slope of the Fe profile is steeper compared to the Virgo cluster \citep{simi15}. The Fe abundance peaks in the core of  Abell 3112 (similar to those of S and Si), however it becomes uniform beyond $0.2R_{200}$ at a level of $\sim0.2-0.24\ Z_{\odot}$. The uniform abundance in the outskirts of the cluster is consistent with the Fe abundance observed in the {\it Suzaku} observations of the nearby clusters, e.g. the Perseus cluster \citep{werner13}. Additionally, the mean observed value at the outskirts of Abell 3112  is consistent with both the Perseus and Virgo clusters. 

\subsection{Radial Distribution of SN Ia to SN cc Fraction}
\label{sec:snapec}
A commonly used method to constrain the distribution of SN enrichment in clusters of galaxies is to examine the relative abundances of metals which are produced by SN Ia and SN cc \citep{deplaa07}. Detailed studies of high signal-to-noise {\it Suzaku} data of nearby Perseus cluster (z =0.018) and Virgo cluster (z= 0.004) have provided tight constraints on the fractional distribution of SN enrichment out to R$_{200}$ using S, Si, and Mg abundance ratios with respect to Fe \citep{werner13, simi15}. However, it is challenging to perform this method for higher redshift clusters, such as Abell 3112 since the detection of abundances of key elements, e.g., Si and S, extends only out to an intermediate radius ($\sim0.5R_{200}$). Additionally, the uncertainty of the observed Mg abundance is large in our observations. Therefore, we use an alternative approach here to measure the SN fraction out to the virial radius.

To investigate the percentage contribution of SN explosions that enrich the ICM, we fit the spectra with the \textit{snapec} model implemented in the {\it XSPEC} fitting package \citep{b12b}.\footnote{\url{http://heasarc.gsfc.nasa.gov/xanadu/xspec/models/snapec.html}} The {\it snapec} model compares the SN yields available in the literature to X-ray spectra in a given energy band. The model has five free parameters: the total integrated number of SNe ($N_{SNe}$) per $10^{12}$ M$_{\odot}$ of ICM plasma - i.e., rescaled to yield values appropriate for cluster cores since the cluster's formation; the ratio of SN Ia to SN cc ($R$); plasma temperature (\textit{kT}), redshift, and normalization. After the fit is performed the goodness of the fit can be used as a test for SN yields. The advantage of this model is that it uses all available elements to constrain the fractional contribution of SNe to chemical enrichment of the ICM as opposed to determining SN enrichment from individual elemental abundance ratios. The {\it snapec} model provides  a self-consistent set of physical parameters, SN fraction and the total number of SNe. Therefore, statistical uncertainties on these parameters are greatly reduced because of the larger number of elemental abundance measurements used in deriving the constraints. Additionally, the method allows the user to choose between different SN enrichment models, and the goodness of the overall fit can be used to test SN enrichment models when finer resolution X-ray observations are available \citep[see][for \textit{Hitomi} simulations]{b12b}. This method is specifically helpful for the case here where we have low signal-to-noise data.

\begin{table}[h!]
\centering
\caption{Best-fit Parameters of the {\it snapec} Model to the {\it Suzaku} Spectrum of Region 1}
{
\normalsize
{\renewcommand{\arraystretch}{1.5}
{\begin{tabular}{lccc}
\hline\hline
SN Ia Model	& $N_{SNe}$  & $R$ & \textit{C}-stat   \\ 
		& ($\times10^{9}$)				& 	& (dof)	\\\hline
W7   & 3.61 $\pm$ 0.16 &  0.10 $\pm$ 0.01 & 1112.4 (840) \\
W70 & 3.59 $\pm$ 0.25 & 0.10 $\pm$ 0.02 & 1108.9 (840)  \\
WDD  & 3.24 $\pm$ 0.10 & 0.12 $\pm$ 0.02 & 1108.1 (840) \\
CDD &  3.18 $\pm$ 0.15 & 0.12 $\pm$ 0.01 &  1108.8 (840) \\
CDDT &  3.08 $\pm$ 0.28 & 0.41 $\pm$  0.09 &  1173.3 (840) \\
ODDT &  3.06 $\pm$ 0.21 & 0.18 $\pm$ 0.03 &  1112.3 (840) \\
\hline
\end{tabular}}}}
\label{table:reg1-snapec}
\end{table}

We use a variety of SN yields from the literature in this work. Among the SN Ia yields are one dimensional spherically symmetric slow deflagration models W7 and W70; delayed detonation models referring to WDD and CDD from \citet{iwa99} and \citet{nomoto06} (I99 and N06 hereafter); and two dimensional delayed detonation models including symmetric (CDDT) and asymmetric (ODDT) explosions from \citet[][M10 hereafter]{maeda10}. Meanwhile, for the SN cc yields, we use the \citet{iwa99} Salpeter-IMF-average yields calculated for a large range of progenitor masses ($10-50 ~M_{\odot}$) and metallicities (0--1 $Z_{\odot}$). We first fit the spectrum extracted from Region 1 using a set of yields from various SN enrichment models. The goodness of the fits of various SN Ia yields are shown in Table \ref{table:reg1-snapec}. We find that I99 WDD model describes the {\it Suzaku} data of the immediate core region the best, with $C$-stat of 1108.13 (840 dof), where we have the highest signal-to-noise data. The I99 W7, CDD, and WDD SN Ia models produce equally good fits to the data with $\Delta$C-stat of 3--4 for the same number of d.o.f. Indeed, W7, WDD, and CDD models predict similar amounts of Fe and Mg. Since the most significantly detected lines in our spectra are Fe-L and  Fe-K  shell lines, they are likely to be responsible for the similar observed $C$-stat values in our fits with W7, CDD, and WDD models. Slight $C$-stat discrepancy between W7 and WDD (and CDD) model can be due to the under-predicted Si abundance in W7 models. However, due to the saddle difference in abundance yields of elements which are available to us in CCD resolution observations, we cannot distinguish between W7, W70, CDD, and WDD models in this work.

 Additionally, we find that M10 CDDT model produces a significantly worse fit to the data ($\Delta C$-stat=65 for 840 dof) compared to the 1D deflagration and detonation SN Ia models. The current {\it Suzaku} CCD observations can already confirm that there is a disagreement between the {\it Suzaku}  observations core region of Abell 3112 and the M10 CDDT model. The underlying reason is the over-predicted Si abundance and under-predicted Mg abundance in the M10 CDDT model ompared to observations in the core region (Region 1). A similar tension in {\it XMM-Newton} observations of a large sample of clusters is reported by \citet{mernier2016}. The authors suggest that the observed discrepancy is due to high Si/Fe ratio requested by the CDDT model  similarly to our conclusion. Lastly, we find a better agreement with the ODDT model and the {\it Suzaku} data as compared to the  M10 CCDT model. This agreement is also noted in \citet{mernier2016}.

\begin{table}[h!]
\centering
\caption{Best-fit Parameters of the {\it snapec} Model Obtained Using I99 WDD SN Ia and I99 SN cc yields}
{
{\renewcommand{\arraystretch}{1.5}
\normalsize
{\begin{tabular}{cccc}
\hline\hline
		& $N_{SNe}$  & $R$ & \textit{C}-stat   \\ 
		& ($\times10^{9}$)			& 	& (dof)	\\\hline
Region 1   & 3.24 $\pm$ 0.10 &  0.12 $\pm$ 0.02 &1108.1 (840) \\
Region 2 & 1.96 $\pm$ 0.36 & 0.16 $\pm$ 0.02 &1079.9 (842)  \\
Region 3  & 1.48 $\pm$ 0.13& 0.12 $\pm$ 0.04 & 1008.9 (850)\\
Region 4  & 1.22 $\pm$ 0.12& 0.13 $\pm$ 0.05 & 337.3 (259) \\
Region 5  &  0.87 $\pm$ 0.17 & 0.11 $\pm$ 0.06 &  244.2 (151) \\
\hline
\end{tabular}}}}
\label{table:snapec}
\end{table}

We note that the goal of the paper is to determine the distribution of SN fraction out to $R_{200}$ rather than individual testing the SN Ia models. Therefore, we use the  I99 WDD SN Ia and I99 SN cc yields providing the best-fit to the highest signal-to-noise data we have here in determining SN fractions. In the {\it snapec} model fits of Region 1 the model parameters $N_{SNe}$, $R$, \textit{kT}, redshift, and normalization are left free. The best-fit parameters of the {\it snapec} models are given in Table \ref{table:snapec}. The temperature measurements between 2T {\it snapec}  (5.29~$\pm$~0.2 keV, 3.24~$\pm$~0.12 keV) and 2T {\it apec}  fits are consistent with each other within the 1$\sigma$ confidence level. The best-fit $N_{SNe}$ and $R$ are respectively (3.24~$\pm$~0.10)~$\times10^{9}$ and 0.12 $\pm$ 0.02. To calculate the total number of SN explosions that enrich the ICM, the parameter $N_{SNe}$ should be rescaled with the projected gas mass within the spectral extraction region \citep[see][for details]{b12b}. The gas mass within 0--2$^{\prime}$ is 3.3~ $\times10^{12}$$M_{\odot}$ \citep[see ][ for mass profiles]{b12a}. Applying a conversion factor of 3.3 (in the units of $10^{12}$~M$_{\odot}$), we find that the ICM of the core of Abell 3112  has been enriched by a total of 1.00 $\pm$ 0.03 $\times10^{9}$ SN explosions within a 12.5 billion year period. This result is consistent with the reported total number of SN explosions (1.06 $\pm$ 0.34 $\times10^{9}$) from {\it XMM-Newton} observations \citep{b12b}.

\begin{figure}[h!]
\centering
\hspace{-4mm}\includegraphics[width=0.49\textwidth]{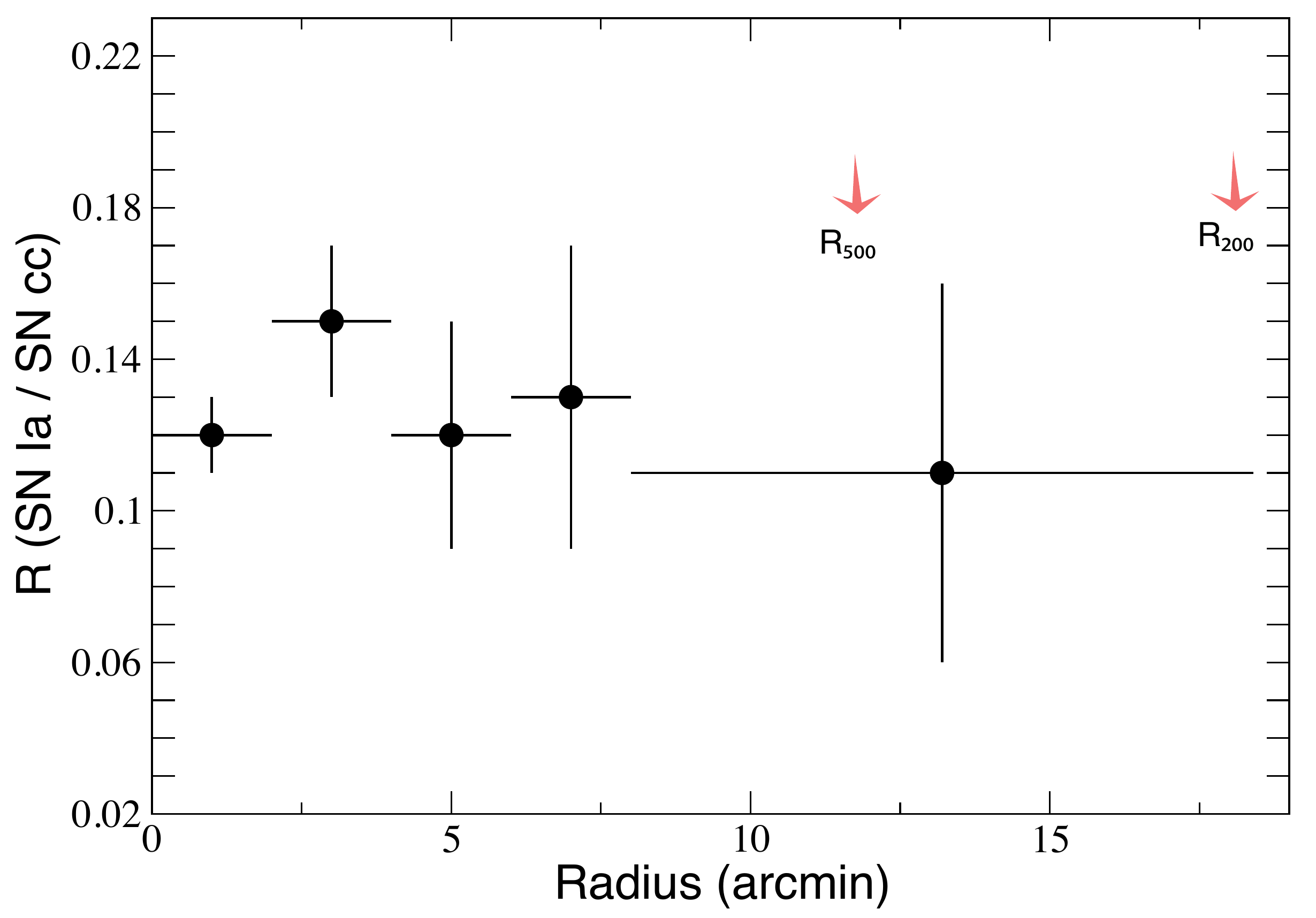}
\caption{The  radial distribution of the SN Ia to SN cc ratio in Abell 3112 out to R$_{200}$ obtained from the fits of X-ray spectra with I99 WDD SN yields. The statistical errors corresponding to $\Delta C=1$ with the systematic uncertainties described in Section \ref{sec:syst} are added to the total error budget shown in the figure. The ratio of the SN Ia to SN cc is fairly uniform from the core to the outskirts. }
\label{fig:snratio}
\end{figure}

The observed fraction $R$ in the {\it Suzaku} observations corresponds to a SN Ia fraction of 11\% in the 0-2$^{\prime}$ region of the cluster. We note that in their results \citet{b12b} report a SN Ia fraction of $\sim$ 30\% in the inner 52~kpc (0.6$^{\prime}$) core region of the cluster. The discrepancy in the SN Ia fraction indicate that SN fraction is diluted by the large PSF size of the \textit{Suzaku} mirrors. The observed difference of the radial profiles of SN Ia products, e.g., S and Si between the {\it XMM-Newton} and {\it Suzaku} observations, indeed, indicates a similar offset.

Using the I99 WDD models in the {\it snapec} fits of Region 2, we find that $N_{SNe}$ is 1.96 $\pm~0.36 \times10^{9}$, while $R$ is 0.16 $\pm$ 0.02 with $C$-stat=1079.9 (for 842 dof). The reported enclosed gas mass is 7$\times10^{12}$~M$_{\odot}$ in Region 2 based on the \citet{bulbul2010} models \citep[see][]{b12b}. Normalizing the $N_{SNe}$ with a factor of 7, we find that the total number of SN explosions enriching the ICM in Region 2 is $(2.8 \pm 0.5) \times10^{8}$ within 12.5 billion years.

The low signal-to-noise data in the spectra of Regions 3, 4, and 5 do not allow us to determine of both $N_{SNe}$ and normalization of the {\it snapec} model simultaneously due to the degeneracy between these variables. The normalization of the {\it snapec} model is essentially determined from the continuum level, and it should be consistent with the normalization parameter of the {\it apec} model. We therefore use the normalization constrained from the {\it apec} model fits for the spectra of Regions 3, 4, and 5  and allow them to vary within their 1$\sigma$ ranges. The best-fit parameters of the {\it snapec} model obtained with this method are shown in Table \ref{table:snapec}. For all the spectra we find acceptable fits to the {\it snapec} model. 

The distribution of  $R$ (=SN Ia/SN cc) is shown in Figure \ref{fig:snratio} from the cluster core  out to R$_{200}$. We note that the systematic uncertainties are included in the error bars shown in the figure. We find that the SN Ia and SN cc ratio, R, is consistent with a uniform SN Ia contribution to the enrichment, with $R\sim0.13$. The SN Ia fraction of 12--16\% (of the total SN explosions) are consistent with the enrichment of the solar neighborhood \citep{tsujimoto95}. This uniformity suggests that both SN Ia and SN cc enrichment of the
ICM outside of the core occurred at an early epoch. Since star formation in galaxy clusters occurs at $z > 2 $ \citep{tran07}, this implies that the SN Ia that enrich the ICM are of the prompt variety, exploding with short delay following this early epoch of star formation. A similar conclusion is inferred from measurements in the low redshift Perseus and Virgo clusters \citep{werner13,simi15}, and the early enrichment timescale is consistent with studies of mass-selected samples of galaxy clusters with redshift $0 < z < 1.5$  \citep{ettori15, mcd16}.

\section{Conclusions}
\label{sec:conc}
In this work we present an analysis of deep {\it Suzaku} (1.2~Ms of total XIS exposure) and {\it Chandra}  (72ks) observations of Abell 3112, to constrain the distribution of SN enrichment of the ICM from the cluster core out to the cluster's virial radius using various published SN yields \citep{iwa99,nomoto06,maeda10}. To constrain the SN fraction we use an {\it XSPEC} model, which is capable of fitting X-ray spectra with pre-defined SN yields from the literature.

Deep {\it Suzaku} observations of this relaxed archetypal cluster allow us to measure the plasma temperature and metal abundance out to cluster's virial radius. We find that temperature constraints from {\it Suzaku} observations are in agreement with previous {\it XMM-Newton} observations within R$_{500}$. The temperature profile peaks around $\sim$~4.7~keV and declines to 3.12~$\pm$~0.70 keV around the virial radius of the cluster. 

We are also able to extend the measurements of metal abundances out to the cluster's virial radius. We find that the metallicity of the ICM is 0.22 $\pm$ 0.08 $Z_{\odot}$ in the outskirts of the cluster near the virial radius and is consistent with the reported  metallicities in nearby clusters \citep{werner13,simi15, bulbul16a}. The observed decline in the Fe abundance is steeper compared to the Mg profile, however, the Fe profile becomes uniform beyond the overdensity radius of $0.2R_{200}$.

We find that the W7, CDD, and WDD SN Ia models produce similar goodness-of-fit to the {\it Suzaku} data. The best-fit SN fraction and the total number of SN parameters obtained from these models are consistent with each other. However, a 2D delayed detonation SN Ia model M10 CDDT produces significantly worse fits to the X-ray spectrum of the central region of the cluster. This suggests that the CDDT models are insufficient to reproduce observed metal abundances (e.g. Si and Mg) in the cores of cluster of galaxies. Nonetheless, accurate testing of SN Ia models using galaxy cluster spectroscopy requires higher spectral resolution. Unfortunately, it will have to wait for the launch of the next calorimeter mission \citep[see][]{b12b,pointecouteau13}.

The distribution of the SN Ia fraction to the total number of SN explosions changes between 12--16\% based on the I99 WDD delayed detonation models, which produce the best-fit to the X-ray spectra. This fraction is consistent with the observed fraction in our Galaxy and proto-solar abundances (15--25\%) \citep{tsujimoto95}. We find that the distribution of the SN Ia fraction is fairly uniform out to the cluster's virial radius. The homogenous SN fraction points to an early (z$\sim$2--3) metal enrichment and mixing originating from an intense period of star formation activity in the cluster outskirts, and it also suggests that the metals are well-mixed into the ICM.  Furthermore, our results are in agreement with the early enrichment timescale inferred from nearby Perseus and Virgo clusters \citep{werner13,simi15} and a mass-selected sample of galaxy clusters reported by \citet{mcd16}.

\section{Acknowledgements}
We are grateful to the referee for the insight and detailed comments which helped to improve the manuscript.
The authors thank Francois Mernier and Tulun Ergin for their valuable comments and suggestions.  E.B. acknowledges support by NASA through contracts NNX14AF78G and NNX123AE77G. E.N.E. would like to thank both  Bogazici University BAP under code 5052 and Tubitak-113F117 for financial support. E.D.M acknowledges support from NASA grants NNX09AV65G and  NNX10AV02G.


\end{document}